# Dual gate control of bulk transport and magnetism in the spin-orbit insulator Sr$_2$IrO$_4$

Chengliang Lu[1, 2*], Shuai Dong[3], Andy Quindeau[1], Daniele Preziosi[1], Ni Hu[1, 4],
& Marin Alexe[1, 5*]

[1] *Max Planck Institute of Microstructure Physics, Weinberg 2, D-06120 Halle (Saale), Germany*

[2] *School of Physics, Huazhong University of Science and Technology, Wuhan 430074, China*

[3] *Department of Physics, Southeast University, Nanjing 211189, China*

[4] *Department of Physics, Hubei University of Technology, Wuhan 430069, China*

[5] *Department of Physics, Warwick University, Coventry CV4 7AL, United Kingdom.*

---

[*] Correspondence should be addressed to C.L.L. (cllu@mpi-halle.mpg.de) and M.A. (M.Alexe@warwick.ac.uk)

Abstracts: The *5d* iridates have been the subject of much recent attention due to the predictions of a large array of novel electronic phases driven by twisting strong spin-orbit coupling and Hubbard correlation. As a prototype, the single layered perovskite $Sr_2IrO_4$ was first revealed to host a $J_{eff}$=1/2 Mott insulating state. In this material, the approximate energy scale of a variety of interactions, involving spin-orbit coupling, magnetic exchange interaction, and the Mott gap, allows close coupling among the corresponding physical excitations, opening the possibility of cross control of the physical properties. Here, we experimentally demonstrate the effective gate control of both the transport and magnetism in a $Sr_2IrO_4$-based field effect transistor using an ionic liquid dielectric. This approach could go beyond the surface-limited field effect seen in conventional transistors, reflecting the unique aspect of the $J_{eff}$=1/2 state. The simultaneous modulation of conduction and magnetism confirms the proposed intimate coupling of charge, orbital, and spin degrees of freedom in this oxide. These phenomena are probably related to an enhanced deviation from the ideal $J_{eff}$=1/2 state due to the gate-promoted conduction. The present work would have important implications in modelling the unusual physics enabled by strong spin-orbit coupling, and provides a new route to explore those emergent quantum phases in iridates.

# I. Introduction

The 5$d$ iridates have emerged as interesting model systems for studying new phases of matter arising from the large relativistic spin-orbit coupling (SOC, $\lambda\sim 0.5$ eV) that is comparable to other relevant parameters such as Hubbard interaction $U$ and electronic bandwidth $W$ [1]. The interplay of these fundamental interactions would lead to the realization of a large array of novel electronic states including the $J_{eff}=1/2$ Mott spin-orbit insulating state [2], Weyl semimetals with Fermi arcs [3], correlated topological insulator [4], Kitaev spin liquid [5], excitonic magnetism of pentavalent $Ir^{5+}$ ($5d^4$) [6], etc. Recently, the $J_{eff}=1/2$ Mott state, a profound manifestation of the strong SOC, was first identified in layered perovskite $Sr_2IrO_4$ (SIO) [2]. In this scenario, the collective action of SOC-driven band-splitting and Coulomb repulsion triggers a Mott transition, resulting in a small charge gap $\Delta\sim 0.3$ eV [7]. Interestingly, the exchange interaction of the $J_{eff}=1/2$ isospins (entangling the spin and orbit momentum) was revealed to have an energy of $J_{ex}\sim 0.1$ eV approximate to the charge gap in SIO, pointing to a possible intimate coupling between charge and isospin excitations [8-11].

While the nature of magnetic coupling between these isospins is still an open issue, recent predictions of high-$T_C$ superconductivity in doped SIO have stimulated great interest in searching for doping-induced unusual quantum phases related to the $J_{eff}=1/2$ state [12-15]. Subsequent experiments indeed revealed an interesting insulator-metal transition, pseudogaps and Fermi arcs in the metal phase in doped SIO [16-21]. Although these observations highly resemble those seen in doped cuprates, the proposed superconducting state has not yet been reported even at the chemical solubility limit of substitution [22], marking the unconventional physics enabled by the strong SOC in the 5$d$ iridates. On one hand, while there is a long list of similarities between SIO and parent compounds of cuprate superconductors such as $La_2CuO_4$, some features of them are still in marked contrast to each other. In $La_2CuO_4$ for instance, the exchange interaction ($J_{ex}\sim 0.1$ eV) is remarkably smaller than the charge gap ($\Delta\sim 2$ eV) [23, 24], basically different from the case of SIO as aforementioned, indicating distinct differences in the interaction of charge and magnetic degrees of freedom between the two materials. On the other hand, the Ir-O-Ir bond angle $\varphi$ was suggested to play a salient role in determining the electronic structure in SIO [9, 25], while the modulation of $\varphi$ is actually a common consequence when tuning carrier density during chemical doping, seriously complicating the related physical picture [16-18, 20, 26]. In addition, a recent optical spectroscopic study argued that short-range antiferromagnetic

(AFM) corrections play an indispensible role for the Mott-metal crossover [20], which is distinct from the classic concept of insulator-metal transition. In this sense, new approaches are urgently desired to clear away those complexities, and clarify the role of carrier-doping in tuning the physical properties of the $J_{eff}$=1/2 state in SIO. This would have important implications in exploring the expected superconductivity in doped SIO, and modeling the 5$d$ iridates involving a strong SOC.

The field effect transistor (FET) is one of the most elegant ways to tune the carrier density in materials without introducing parasitic effects such as quenching disorder. The newly developed electric-double-layer transistor (EDLT) technique involving an organic ionic liquid is particularly powerful owing to its very high accumulation of carrier density up to ~$10^{15}$ cm$^{-2}$ [27-30]. For example, abnormal gating effects were recently evidenced in a few newly fabricated Mott-insulator EDLTs, resulting from electric control of inherent collective interactions of the channel material [30-33]. As stated before, SIO shares similar Mott physics with those strongly correlated 3$d$ electron oxides, but its unique physics involving a large SOC builds a subtle energy balance in this material. A very recent study reported unusual Zeeman-type spin splitting related to strong SOC in a WSe$_2$-EDLT [34]. Therefore, it would be of interest to find exotic gating effects in spin-orbital insulator SIO-based EDLTs. In this work, an unusual bulk gating effect on the transport was identified in micro-patterned SIO-ELDTs. This effect, which couples strongly to the magnetism of SIO, is fundamentally different from the surface-limited operation seen in conventional FETs [31, 35]. This dual electric-control of three dimensional conduction and magnetism distinguishes the unique physics of the $J_{eff}$=1/2 state in SIO, and could pave the way for an active control of electronic structure in the 5$d$ iridates.

## II. Experiments

Epitaxial SIO films with various thicknesses ranging from 10 nm to 40 nm were grown on STO (001) substrates with vicinal surface using pulsed laser deposition (PLD) [36]. X-ray diffraction (XRD) analyses were carried out using a Philips X'Pert diffractometer. Laterally gated devices were fabricated by standard photolithography and argon ion etching. The dimensions of a channel were 50 μm in width and 200 μm in length. Gold electrodes were deposited by thermal evaporation. An organic ionic liquid N, N-diethy1-N-(2-methoxyethyl)-N-methylammonium bis(trifluoromethylsulfonyl)imide (DEME-TFSI) was used for the gating experiments. The gate voltage was applied at $T$=220 K for 0.5 hour, after which the sample was cooled down to 180 K to freeze the electrolyte in a glassy state. After an additional equilibration for 15 min, the transport measurements were performed using a standard four-probe method in a physical property measurement system (PPMS, Quantum Design). All magneto-transport measurements were performed using the same four-probe method.

### III. Results

$Sr_2IrO_4$ has a tetragonal crystal structure and an isospin canting-induced weak ferromagnetic (FM) ground state with $T_C$~230 K (see Appendix) [37, 38]. To best preserve those bulk properties and thus simplifying our model system, SIO thin films were grown on (001) $SrTiO_3$ (STO) substrates because of their good lattice match [37, 39]. Fig. 1(a) shows a XRD $\theta$-$2\theta$ scan pattern around the (0012) diffraction peak of a 20-nm SIO thin film, indicating the high crystal quality of the film. A coherent epitaxial growth of the film was further confirmed by reciprocal space mapping shown in Fig. 1(b). Devices for gating experiments (Fig. 1(d)) were then fabricated from 20-nm SIO thin films.

As shown in Fig. 1(c), upon sweeping the gate voltage $V_G$ at 220 K, the source-drain current $I_{SD}$ ($V_{SD}$=50 mV) shows an abrupt increase at $V_G$<-0.5 V, demonstrating a $p$-type FET characteristic. The hole carrier accumulation process is schematically shown in Fig. 1(e). Regarding the hysteresis in the transfer curve, the gate-induced formation of oxygen vacancies was usually cited as an important origin in EDLTs [32]. However, here the observed $p$-type FET characteristic (negative $V_G$ charging) can basically rule out such a mechanism. Another possibly involved factor is the chemisorption of OH$^-$ on the channel surface due to the humidity in the ionic liquid [40]. Such an effect should not be substantial considering the repeatability and the negligible leakage current $I_G$<1 nA. Therefore, we would rather argue an intrinsic modulation of the electronic structure being responsible for the hysteresis behavior upon gating, as further evidenced in the following.

After introducing carriers into the channel by applying different $V_G$, the temperature ($T$) dependence of the four-terminal sheet resistance $R_s$ was measured, as shown in Fig. 2(a). First, the sheet resistance $R_s$ is significantly reduced by $V_G$. However, despite a reduction in $R_s$ larger than two orders of magnitude at low $T$, semiconducting-like behavior of transport persists for all cases in the entire temperature range. By performing Hall-effect measurements at 180 K, the sheet carrier density $n_s$ and carrier mobility $\mu_s$ were evaluated and plotted as a function of $V_G$ in Fig. 2(b). In the non-gated case ($V_G=0$) the SIO film shows a *p*-type characteristic with $n_s=1.7\times10^{14}$ /cm$^2$, corresponding to a volume carrier density of $n=8.3\times10^{19}$ /cm$^3$ close to the bulk SIO [41]. This agrees with the *p*-type electric field operation seen above. With decreasing $V_G$ to -2 V, $n_s$ increases drastically by 20 times to $3.8\times10^{15}$ /cm$^2$, marking the accumulation of hole carriers of very high density in SIO. We note that a recent paper shown nearly the same gate modulation of the charge carrier density [42]. In contrast to the massive increase in $n_s$, the carrier mobility $\mu_s$ shows a clear decrease at $V_G<-1$ V. Thus, the reduction of $R_s$ is solely due to the accumulation of hole carriers of high density.

Here, one more interesting feature is worthy mentioning, *viz.* the carrier density $n_s$ growing with increasing $V_G$ nonlinearly. This is quite different from the conventional electrostatic surface charging process seen in most EDLTs (or FETs), where $n_s$ linearly increases with $V_G$ following the classic capacitor model ($Q=CV_G$, $Q$ is the total accumulated charge and $C$ is the capacitance of the electrolyte). To capture more details about this discrepancy, a straightforward pathway is to compare the experimentally obtained $n_s$ with the nominal $n_s$ calculated from the classic capacitor model. For example, in the present case at $V_G=-2$ V, assuming a typical $C$ value of an EDL of 10 μF/cm$^2$ [30], the calculated nominal $n_s$ is $\sim1.3\times10^{14}$ /cm$^2$, which is $\sim$30 times lower than the obtained $n_s$ ($3.8\times10^{15}$ /cm$^2$) from the Hall-effect measurements.

This much higher $n_s$ together with the nonlinear $n_s$-$V_G$ dependence clearly demonstrates the emergence of massive additional hole carriers beyond the simple electrostatic surface charge accumulation. To check the possible source of such remarkable increase in carrier concentration, we further analyzed the thickness (*t*) dependence of the gating experiments. As shown in Fig. 2(c), similar $I_{SD}$-$V_G$ curves with hysteresis can be seen for all films with different thickness *t*. Obviously, the maximum $I_{SD}$ around $V_G=-1.2$ V increases with *t*, and a linear relationship is obtained after plotting the conductance $\sigma_s$ as a function of *t* (Fig. 2(d)), giving rise to a constant conductivity of 6.5 S/cm at 220 K. This feature suggests a bulk gating effect on the transport in the present SIO-EDLT, similar to the behavior seen in VO$_2$-EDLT [30] but essentially different from that in SmCoO$_3$-EDLT [31].

As revealed by theoretical calculations, enhanced itinerancy could drive the real ground state away from the ideal $J_{eff}=1/2$ state, thus modifying the electronic structure and the magnetism [43]. Therefore, the promoted conduction associated with a very high density of the hole carriers allows us to anticipate a remarkable impact on the special magnetism in the present SIO-EDLT. Direct measurements of magnetization are extremely challenging because of the inherent antiferromagnetic structure and the very small size of the device. We then turn to measure the magnetoresistance (defined as MR=[$R(H)$-$R(0)$]/$R(0)$, where $H$ denotes the applied magnetic field), which is also a direct signature of magnetism [34]. As shown in Fig. 3(a), the MR of the non-gated film ($V_G$=0) shows a sudden drop at $H$~0.2 T, owing to the reduction of magnetic scattering associated with an $H$-induced isospin-flip transition [1, 17, 39]. The isospin-flip transition is illustrated Fig. 3(b), which can be evidenced by a direct comparison between magnetization and transport (see Appendix). However, after applying $V_G$=-2 V, the isospin-flip transition induced drop in MR seen in the non-gated film is greatly suppressed, and instead only a weak negative MR can be captured (Fig. 3(d)). This indicates that the isospin-flipping process still occurs but with a much smaller critical field and net FM moment, giving rise to a significantly reduced spin-dependent scattering effect responsible for the observed weak negative MR. Such isospin-flipping process can be well recovered after gating experiments, proving the intrinsic feature of the present results (see Appendix).

Increasing $T$ causes an evident change in MR of the non-gated film shown in Fig. 3(c). First, the magnitude of the MR decreases with increasing $T$, consistent with a weakened FM component due to the large thermal fluctuations at high $T$. Second, an intriguing reversal in MR with clear hysteresis emerges at high $H$, and it becomes more striking with increasing $T$. This certainly doesn't follow the spin-dependent scattering mechanism observed at low $T$ and low $H$. The distinct hysteresis of this reversal and the strong SOC in SIO allow us to propose a possible magnetoelastic coupling effect to be responsible for such a phenomenon. While this intriguing feature of reversal in MR deserves further investigation, a two-component scenario of the MR could be qualitatively concluded from the data, including the low-$T$ negative MR *via* the spin-dependent scattering mechanism, and a high-$T$ positive component possibly associated with a magnetoelastic effect. The two contributions coexist and compete with each other. A similar phenomenon was also observed in other SIO thin films[44], which well matches the above two-component scenario. Therefore, the emergence of the positive MR (or the reversal in MR) could be regarded as a signature of a weakened FM component in SIO thin films. After applying $V_G$=-2 V shown in Fig. 3(d), the MR curve even shows a reversal at $T$=35 K, which is much lower than that of the non-gated film, indicating the suppression of the FM phase upon gating. With increasing $T$ at $V_G$=-2 V, no significant variation can be seen in the MR curve, differing from the non-gated case.

The modulation of magnetism by gating was further confirmed by measuring the out-of-plane anisotropic magnetoresistance (defined as AMR=$[R(\Phi)-R(0)]/R(0)$, where $\Phi$ is the angle between $H$ and the $c$ axis, and $R(\Phi)$ is the resistance when $H$ is applied at $\Phi$) with a perpendicular configuration of source current $I$ and $H$ at various temperatures. The AMR measurement setup is schematically shown in Fig. 4(a). To first capture the detailed correlation between the AMR and the weak FM phase in SIO thin films, we measured the $T$-dependence of magnetization and AMR for a bare SIO thin film, as provided in the Appendix. At low-$T$, the AMR shows a two-fold periodicity with a minimum value at $\Phi$=90º, which can be well described by the low-$T$-enhanced in-plane FM component that gives rise to a larger in-plane spin-dependent scattering effect than out-of-plane. At high-$T$ (~220 K) close to $T_C$=225 K, where the FM component is extremely weak, although the AMR is still two-fold, it becomes positive and the minimum value is switched to $\Phi$=0º. This is beyond the spin-dependent scattering mechanism, and has the same magnetoelastic origin as the feature of reversal in the normal in-plane MR shown in Fig. 3(c). Such switching behavior of the AMR is bridged by the four-fold AMR symmetry in the middle-$T$ region (such as $T$=170 K), composed of both the low-$T$ AMR with a minimum value at $\Phi$=90º and the high-$T$ AMR with a minimum value at $\Phi$=0º, where $M$ shows a quick decrease with increasing $T$. In Ref. [44], similar AMR switching with $T$ was evidenced in a SIO/STO thin film, but the onset temperature (~150 K) is lower than the present value (~220 K), because of the lower $T_C$~190 K. Such direct correspondence between the AMR symmetry and the weak FM phase provides another effective measure of the gate modulation of magnetism in the present SIO-EDLT. For the non-gated film ($V_G$=0) shown in Fig. 4(b), as expected, a two-fold AMR with a minimum value at $\Phi$=90º associated with a relatively large FM component is indeed seen at $T$=35 K (deep in the ground state), and a four-fold AMR related to the thermally weakened FM phase arises at $T$=180 K. However, after applying $V_G$=-2 V, a four-fold AMR symmetry emerges even at $T$=35 K (Fig. 4(c)) where the non-gated film shows a strong two-fold character, confirming the gate suppression of the weak FM phase in the present SIO-EDLT. Moreover, the film under $V_G$= -2 V already shows a switched two-fold AMR with a minimum value at $\Phi =0º$ at $T=180$ K, where the non-gated film still has a four-fold AMR, likely implying a lower $T_C$ in the gated film.

## IV. Discussion and conclusion

Thus far we have demonstrated the concurrent gate control of transport and magnetism in SIO-EDLT, pointing to a close correlation between charge carriers and magnetism in SIO. Interestingly, the as-generated carrier density is much higher than the value calculated following a classic capacitor model. Further thickness dependent gating experiments reveal that the emergent massive carriers are distributed throughout the film, which is beyond a surface limited operation seen conventional FETs. The results seem to point to some kind of electric field induced phase transition in the SIO thin film that is responsible for the hysteretic behavior of the transport and magnetic properties.

Recently, a similar bulk gating effect on transport was first reported in $VO_2$ based EDLT, and structural transformation (from monoclinic to tetragonal) due to oxygen migration away from the film was demonstrated to be the origin [30, 32]. In the present SIO-EDLT, the oxygen migration mechanism can be clearly excluded by the observed $p$-type field effect. Actually, such strong ion motion as well as the consequent structural transition would also not be expected for $Sr^{2+}$ and $Ir^{4+}$, since the two cations are much harder than oxygen to be removed from the SIO film [16]. In addition, the SIO shows pure tetragonal phase up to 600 K [38], which surely doesn't favor a structural phase transition similar to that in $VO_2$-EDLT. However, the strong electron-lattice coupling in SIO would still allow possible lattice modulation upon gating in the present SIO-EDLT with massively enhanced carrier concentration. And the required lattice change could be tiny considering the delicate balance of competitive interactions in SIO. For example, it was reported that its resistivity could increase by more than four orders of magnitude with decreasing $T$ from 300 to 10 K, although the cell volume of SIO just shows a minuscule change (<0.5%) [37].

Comparing with the possible tiny lattice modulation related to electron-lattice coupling, we would rather argue a significant role of modified electronic structure, owing to the remarkably increased carrier density, to be responsible for the gating effect seen in the present work. It is known that the essential ingredients for the novel $J_{eff}=1/2$ Mott insulator is three-fold [2]: i) The strong SOC; ii) A moderate Hubbard $U$; iii) Five electrons on the Ir $5d$ orbitals. This SOC-assisted Mott-type band structure is not rigid but quite sensitive upon the change of carrier density, similar to (probably even more pronounced than) other traditional Mott insulators [45]. Our experiment on the gate-enhanced bulk-like conduction supports this scenario. Upon the gating voltage, a small number of screening charges, accumulated near the interface, violate the above condition iii), which triggers the reduction of the effective Coulomb repulsion energy [45]. Then the effects of both conditions i) and ii) will be modulated because of the subtle balance of the multiple fundamental interactions, and the $J_{eff}=1/2$ state might be modified. Such a chain-reaction-like process is a positive feedback, which will spread within the whole film and release massive charge carriers.

A recent theoretical work, which gave quantitative descriptions of the $J_{eff}=1/2$ state and magnetism in SIO and $Sr_3Ir_2O_7$, indeed revealed close correlation between the $J_{eff}=1/2$ state and itinerancy [43]. According to this model, enhanced itinerancy could drive the ground state away from the ideal $J_{eff}=1/2$ state, and thus suppress the magnetism within the system, which can well explain the different magnetism in SIO and $Sr_3Ir_2O_7$ [17, 43, 46]. The remarkably increased carrier density and concurrent suppression of the weak FM phase evidenced in the present work are consistent with such scenario. Nevertheless, thorough microstructure investigations are desired to clarify the dual gate tuning of transport and magnetism observed in the present SIO-EDLT.

In summary, we experimentally evidence dual gate-control of bulk transport and magnetism in $Sr_2IrO_4$ by utilizing field effect transistor principle. The observed strong coupling between transport and magnetism suggests a salient role of charge carriers in mediating the interaction between the novel $J_{eff}=1/2$ moments. Insulating transport is revealed to exist even at a relatively high hole-doping level (0.19 hole/Ir), which would provide useful parameters for modelling novel phases in doped $Sr_2IrO_4$. Our work has shown a promising route to manipulate the physical properties related to the novel $J_{eff}=1/2$ state in $Sr_2IrO_4$. This opens new ways to experimentally realize the predicted exotic electronic states driven by strong spin-orbit coupling especially in iridates.

**Acknowledgements**

We acknowledge fruitful discussions with Prof. Dietrich Hesse, Prof. Stuart Parkin, Prof. Jun-Ming Liu, and Dr. Xavi Marti. We would like to thank Dr. Ignasi Fina and Mr. Xiang Li for the help with magnetization measurements and discussions. This work was partly funded by DFG via SFB 762. C.L.L. acknowledges the funding from the Alexander von Humboldt Foundation and the National Natural Science Foundation of China (Grant Nos. 11104090 and 11374112). S.D. was supported by the National Natural Science Foundation of China (Grant Nos. 51322206 and 11274060). N.H. was supported by the National Natural Science Foundation of China (Grant No. 11304091). M.A. acknowledges the Wolfson research merit award of the Royal Society.

**Figure captions:**

Figure 1. (a) X-ray diffraction $\theta$-$2\theta$ scan of a film with thickness of 20 nm. (b) The corresponding reciprocal space mapping image around the (109) plane of the film. The subscription "S" and "F" denote the substrate and film, respectively. (c) Source-Drain (S-D) current $I_{SD}$ versus gate voltage $V_G$ transfer curves of the 20 nm thick $Sr_2IrO_4$ electric-double-layer transistor at 220 K. A constant $V_{SD}$=50 mV was used, and the $V_G$ sweep rate was 5 mV/s. The leakage current $I_G$ is in blue, which is less than 1 nA. (d) Sketch of the device structure, in which the orange bar at the center denotes the channel. (e) Sketch of hole-carrier accumulation in the channel upon gating.

Figure 2. (a) Sheet resistances as a function of temperature under various gate voltages. (b) $V_G$ dependence of sheet carrier density $n_s$ and carrier mobility $\mu_s$ obtained by Hall-effect measurements at 180 K. The black open circles denote the nominal $n_s$ calculated from $Q=CV_G$ by assuming a typical $C$ value of 10 µF/cm². (c) Consecutive cycling of transfer curves of 10 nm and 40 nm thick films at 220 K. (d) Sheet conductance $\sigma_s$ derived from the transfer curves as a function of thickness.

Figure 3. (a) Initial magnetoresistance curves after zero magnetic field cooling under various $V_G$. The critical field for isospin flipping is indicated by a vertical dashed line. (b) The isospin-flip transition is illustrated. Measured magnetoresistance at various temperatures for (c) $V_G$ =0, and (d) $V_G$ =-2 V. In (c) and (d), the red (blue) curves show the down- (up-) $H$ sweeps ($H$ parallel to the [100] direction) sweeps.

Figure 4. (a) Setup of anisotropic magnetoresistance measurements. The applied magnetic field $H$=9 T is always perpendicular to the current $I$, and the polar angle $\Phi$ is defined as the angle between $H$ and the [001] direction. Measured anisotropic magnetoresistance (AMR%=[$R(\Phi)$-$R(0)$]/$R(0)$) at various temperatures for (b) $V_G$ =0, and (c) $V_G$ =-2 V. In (b) and (c), the red (blue) curves show the up- (down-) $\Phi$ sweeps.

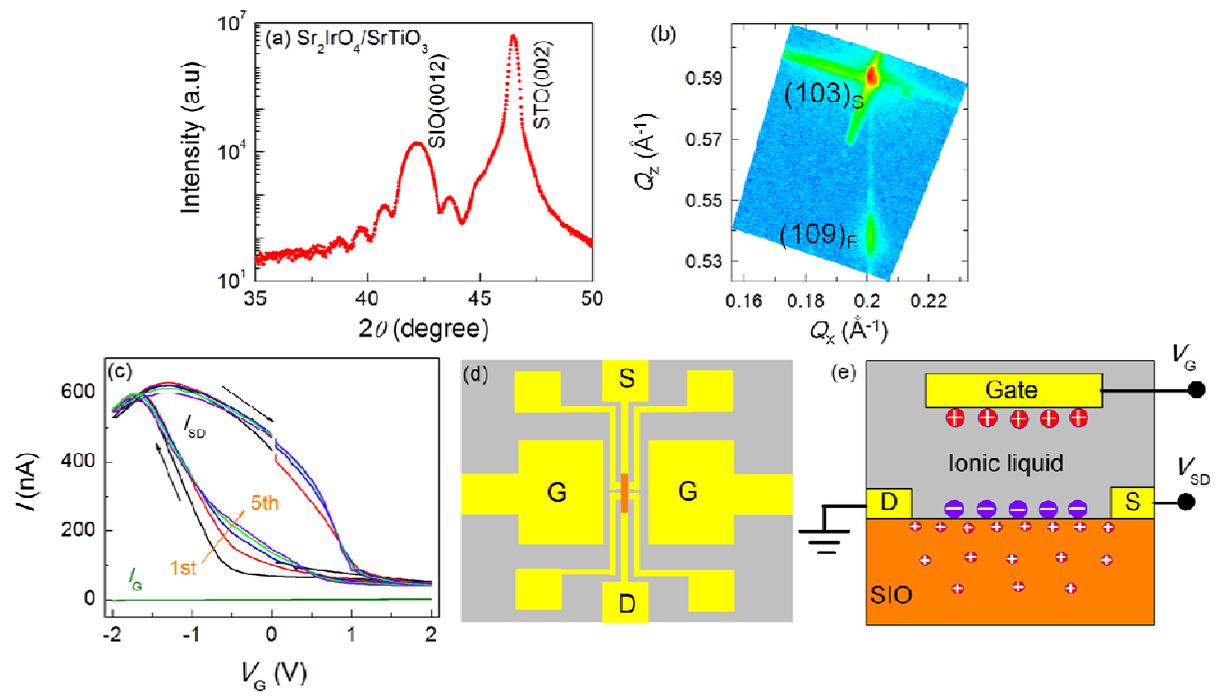

Figure 1

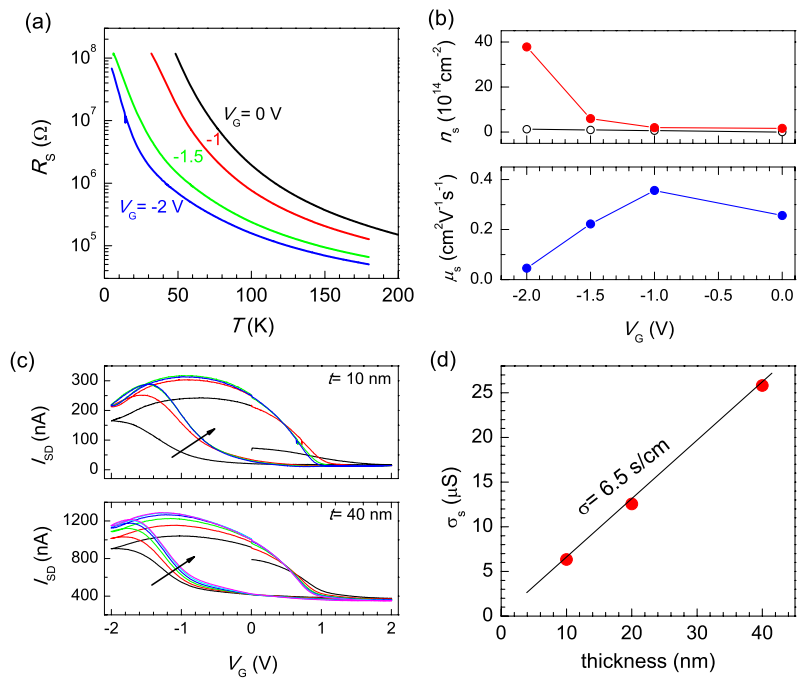

Figure 2

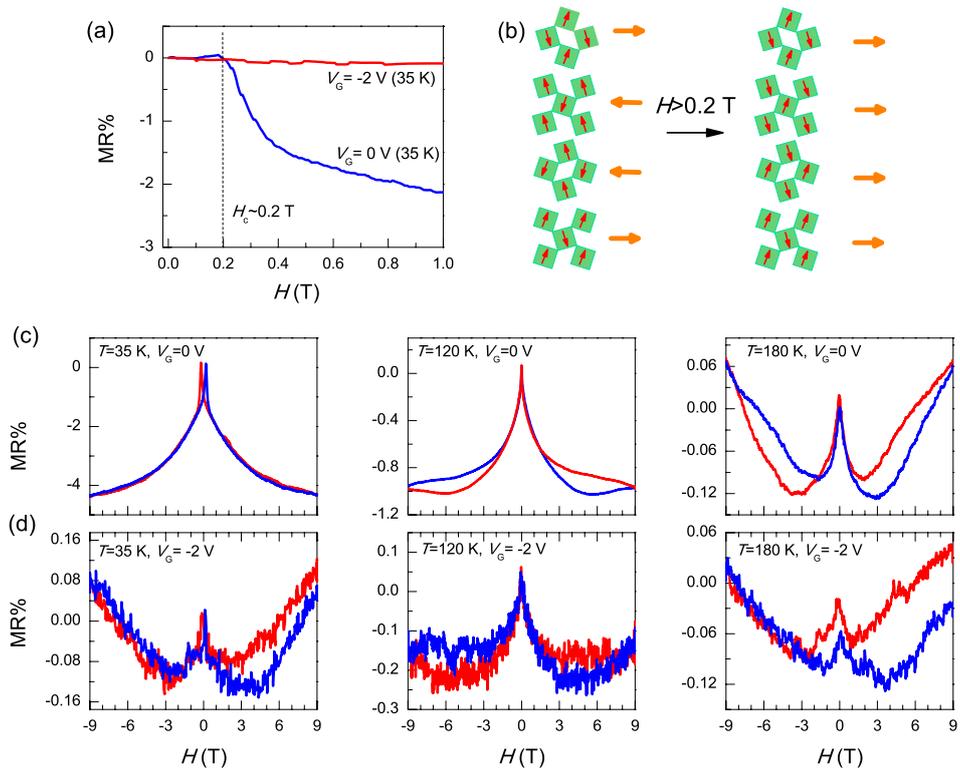

Figure 3

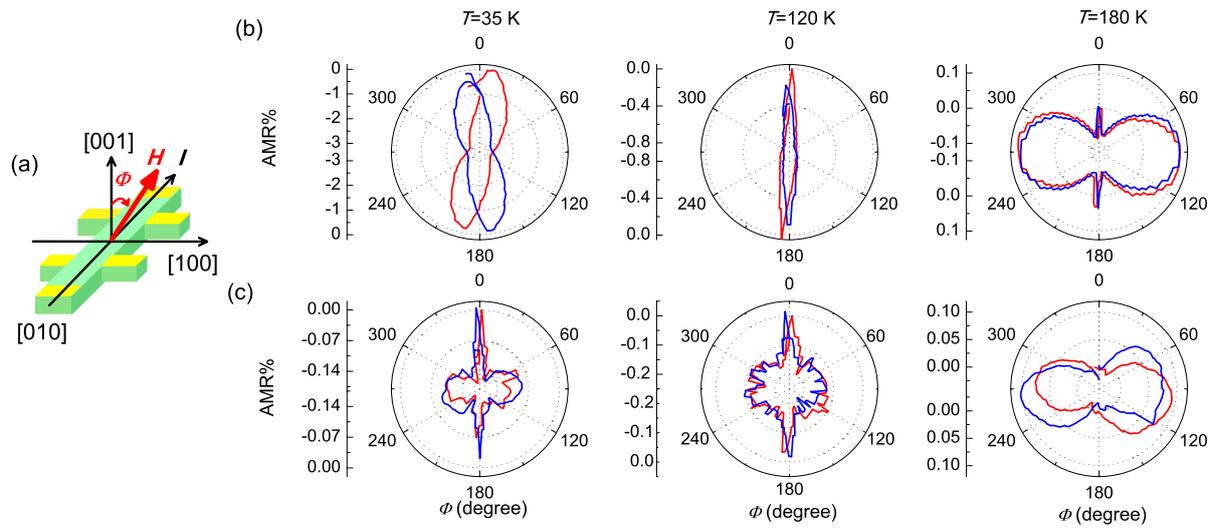

Figure 4

**APPENDIX**

Single layer perovskite $Sr_2IrO_4$ has a tetragonal structure. Each $IrO_6$ octahedron rotates by 11.8º with respect to the *c* axis. Neutron scattering experiments revealed that $Sr_2IrO_4$ has an in-plane antiferromagnetic configuration with Ir isospins deviating 13º from the *a* axis. This isospin canting rigidly tracks the octahedral rotation and gives rise to a net moment in each $IrO_2$ layer [38]. The crystalline and magnetic structure is illustrated in Fig. 5.

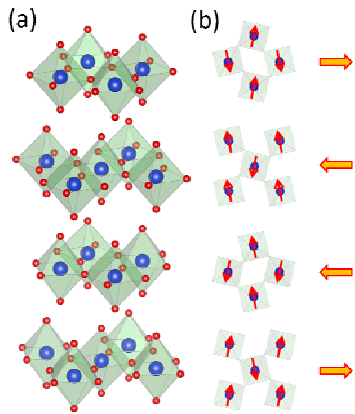

Figure 5. (a) Crystal structure of $Sr_2IrO_4$ consisting of $IrO_2$ layers separated by Sr planes (not shown). The blue and red balls represent Ir and O atoms, respectively. (b) In-plane canted antiferromagnetic structure of isospins (red arrows). The orange arrows denote the net moment of the $IrO_2$ planes.

The application of magnetic field along the in-plane in $Sr_2IrO_4$ could trigger an isospin-flip transition, giving rise to a weak ferromagnetic phase. Consequently, the resistance of the material could show a sudden drop due to the reduction of magnetic scattering. This effect can be directly evidenced by a comparison between magnetization and transport data. In Fig. 6, a clear metamagnetic transition can be seen around 0.2 T, which is accompanied by a drop in resistance.

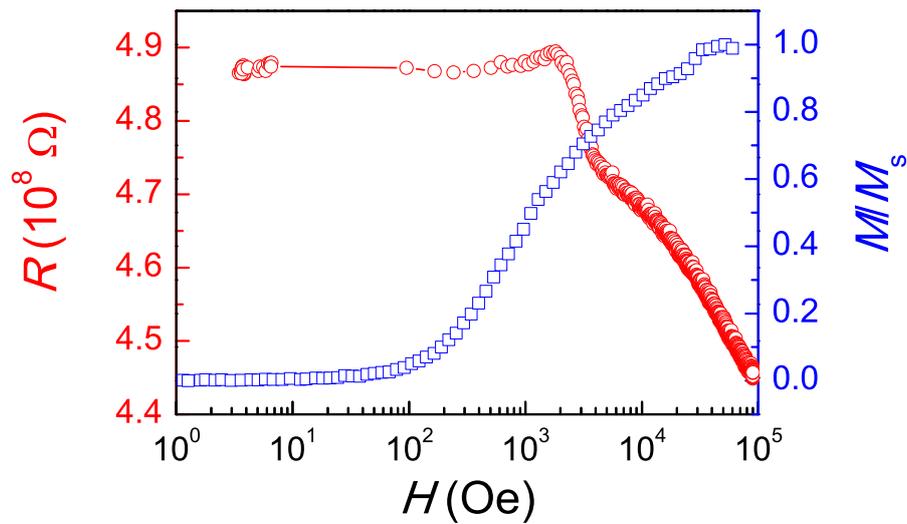

Figure 6. Resistance (red circles) and normalized magnetization ($M/M_s$, blue squares) as a function of magnetic field of a 20-nm $Sr_2IrO_4$/$SrTiO_3$ (001) thin film measured at 35 K.

After the gating experiments, the pristine state seen in the non-gated film can be well recovered. As shown in Fig. 7, the measured magneto-transport after the gating experiments highly resembles those observed in the non-gated film, although there is little difference in the magnitude of MR and AMR. This suggests the intrinsic feature of the present gating experiments.

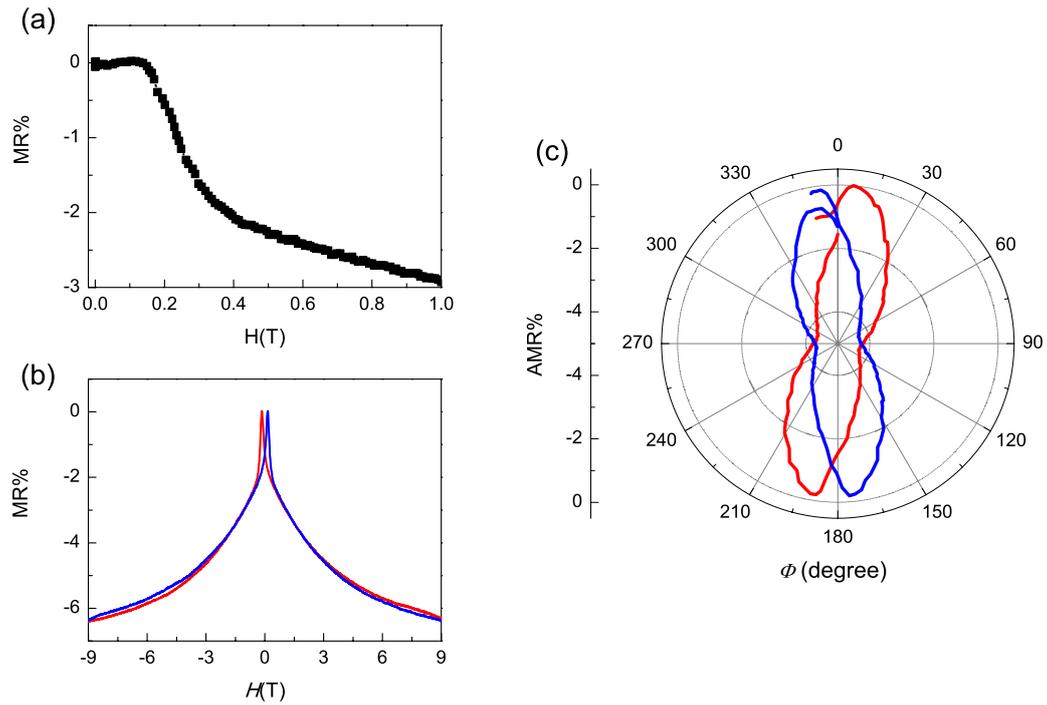

Figure 7. Magneto-transport of the 20 nm $Sr_2IrO_4$ based electric double layer transistor after gating experiments.

Anisotropic magnetoresistance shows close correlation with the weak ferromagnetic phase arising form isospin canting in $Sr_2IrO_4$. In Fig. 8, three clear anomalies (indicated by black arrows) can be seen in the temperature dependence of magnetization curves, which shows one-to-one correspondence to the multiple magnetic transitions in $Sr_2IrO_4$ bulk crystals [17], consistent with optical spectroscopy measurements according to which $Sr_2IrO_4$ thin films grown on $SrTiO_3$ (001) have the same magnetic structure as the bulk counterpart [39]. The measured anisotropic magnetoresistance shows a switching behavior around $T_C$, corresponding to the evolution of the weak ferromagnetic phase with temperature. This close correlation between the AMR symmetry and the weak ferromagnetic phase provides an efficient measure of the magnetic property of the present $Sr_2IrO_4$ field effect devices using an ionic liquid dielectric.

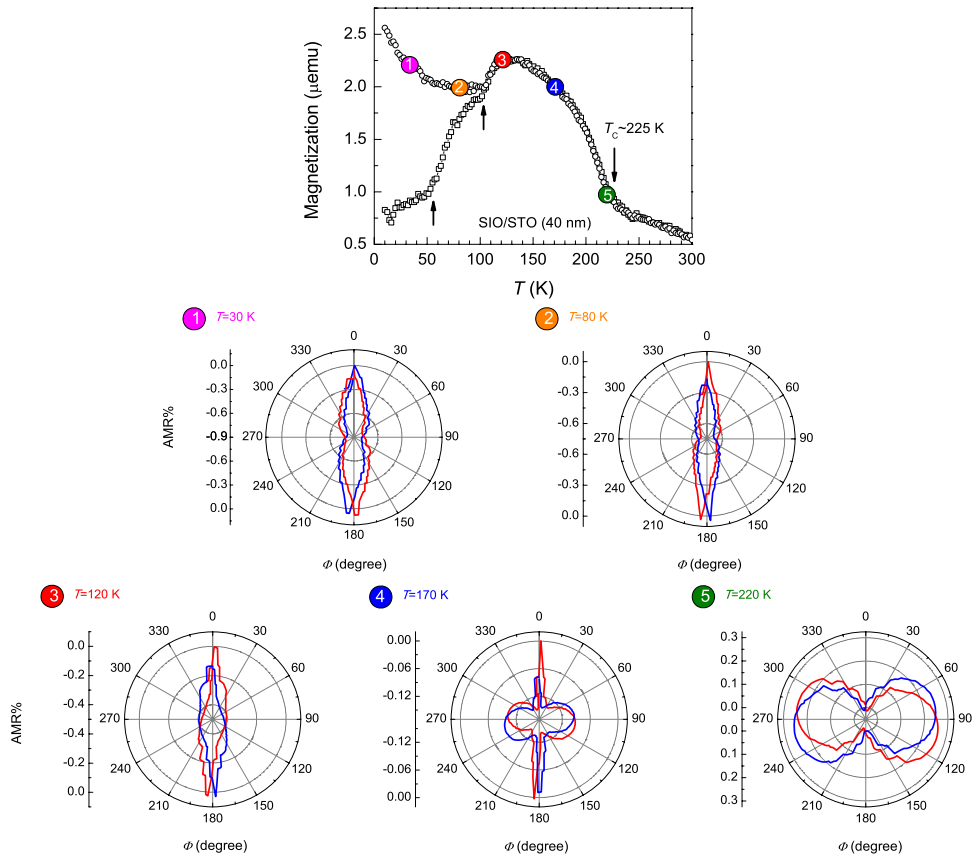

Figure 8. (top) Temperature dependence of magnetization of a 40-nm $Sr_2IrO_4$ thin film. (bottom) Anisotropic magnetoresistance measured at 30 K, 80 K, 120 K, 170 K, and 220 K.